\documentclass[prl,twocolumn,showpacs,floatfix,longbibliography]{revtex4-1}
\usepackage{mathrsfs}
\usepackage{amssymb, amsbsy, amsmath, latexsym, dsfont, array, layout,mathrsfs,color,ulem,bm}
\usepackage[colorlinks,linkcolor=blue,anchorcolor=red,citecolor=blue]{hyperref}

\newcommand{\one}{\mathds{1}}
\newcommand{\ket}[1]{\left|{#1}\right\rangle}
\newcommand{\bra}[1]{\left\langle{#1}\right|}

\newcommand{\ketbrad}[1]{\left|{#1}\rangle\!\langle{#1}\right|}

\usepackage{graphicx}
\usepackage{xspace, stmaryrd, ulem}
\definecolor{delete}{rgb}{1.0, 0.0, 0.0}
\definecolor{edit}{rgb}{0.0, 0.0, 0.9}
\definecolor{comment}{rgb}{0.9, 0.0, 0.0}

\begin{document}

\title{Simulating Exceptional Non-Hermitian Metals with Single-Photon Interferometry}

\author{Kunkun Wang\textsuperscript{1,5}}
\author{Lei Xiao\textsuperscript{1}}
\author{Jan Carl Budich\textsuperscript{2}}\email{jan.budich@tu-dresden.de}
\author{Wei Yi\textsuperscript{3,4}}\email{wyiz@ustc.edu.cn}
\author{Peng Xue\textsuperscript{1}}\email{gnep.eux@gmail.com}

\affiliation{\textsuperscript{1}Beijing Computational Science Research Center, Beijing 100084, China}
\affiliation{\textsuperscript{2}Institute of Theoretical Physics${\rm ,}$ Technische Universit\"{a}t Dresden and W\"{u}rzburg-Dresden Cluster of Excellence ct.qmat${\rm ,}$ 01062 Dresden${\rm ,}$ Germany}
\affiliation{\textsuperscript{3}CAS Key Laboratory of Quantum Information, University of Science and Technology of China, Hefei 230026, China}
\affiliation{\textsuperscript{4}CAS Center For Excellence in Quantum Information and Quantum Physics, Hefei 230026, China}
\affiliation{\textsuperscript{5}School of Physics and Material Science, Anhui University, Hefei 230601, China}

\begin{abstract}
We experimentally simulate in a photonic setting non-Hermitian (NH) metals characterized by the topological properties of their nodal band structures. Implementing nonunitary time evolution in reciprocal space followed by interferometric measurements, we probe the complex eigenenergies of the corresponding NH Bloch Hamiltonians, and study in detail the topology of their exceptional lines (ELs), the NH counterpart of nodal lines in Hermitian systems. We focus on two distinct types of NH metals: two-dimensional systems with symmetry-protected ELs, and three-dimensional systems possessing symmetry-independent topological ELs in the form of knots. While both types feature open Fermi surfaces, we experimentally observe their distinctions by analyzing the impact of symmetry-breaking perturbations on the topology of ELs.
\end{abstract}

\maketitle

{\it Introduction.---}
Phases of quantum matter characterized by topologically robust nodal band structures, such as Weyl semimetals exhibiting remarkable transport properties~\cite{AMV18}, have been a major focus of both theoretical and experimental study for the past decade~\cite{AMV18,HK10,QZ11,WBB14,OHP19,BBK19}. Recently, it has become clear that non-Hermiticity, a common element in open, dissipative systems~\cite{BB98,BBJ02,B07,EMK18}, can qualitatively modify key features of band topology (see Ref.~\cite{BBK19} for a review). This leads to a variety of fascinating phenomena, including the emergence of anomalous topological edge states~\cite{Lee16,XZB17,KEBB18,Xi18,YW18,XDW20,OKS20,ZLL+21}, and the presence of exceptional lines (ELs)~\cite{ZDZ+19}, the non-Hermitian (NH) generalization of nodal lines along which the NH Hamiltonian is nondiagonalizable~\cite{XWD17,MG+18,CB18,BCKB19,CSBB19,YH19,WRZ19,KBS19,YCF20,LSH+19,BN+20}. In particular, at least two distinct classes of NH metals have been theoretically identified, one with ELs protected by NH symmetries that reduce the codimension of exceptional points (EPs)~\cite{BCKB19,Yoshida2019}, the other featuring nodal band structures and knotted ELs with intrinsic, symmetry-independent topology~\cite{CSBB19}. These NH metals are in sharp contrast to their Hermitian, semimetal counterparts, where ELs and Fermi surfaces are reduced to isolated Weyl points and surface Fermi arcs, respectively~\cite{FWD16,BYL17,Ezawa17,TM18,ZPY18}. While nodal band structures in both Hermitian and NH settings have seen a great surge of experimental interest recently~\cite{BWR+16,YLY18,SHN19,ZLL19,XYQ20,CLP20}, a systematic study of unconventional NH metals and their topological stability is still lacking.

\begin{figure}[b!]
\includegraphics[width=0.45\textwidth]{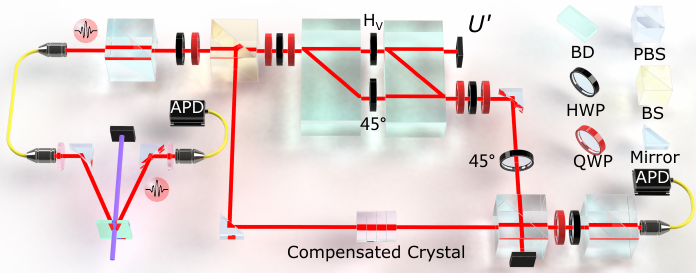}
\caption{Experimental setup. The polarization of the signal photon generated by the standard spontaneous parametric down-conversion is projected into the polarization state $\ket{\psi_\pm}$ with a polarizing beam splitter (PBS), a half-wave plate (HWP) and a quarter-wave plate (QWP), and then goes through the interferometric network.
After passing through a $50:50$ beam splitter (BS), the photon is either transmitted or reflected, and thus separated into different paths. Subsequently, a nonunitary operation $U'$ realized via sets of wave plates and two beam displacers (BDs) is performed on its polarization state in the transmitted mode which acquires a complex phase shift, corresponding to the eigenenergies of $H'$. Finally, the photon is detected by avalanche photodiodes (APDs) resulting in a ``click click'' in coincidence detection event involving another trigger photon. We measure the eigenenergies in momentum space via interferometric measurements.}
\label{fig:setup}
\end{figure}

In this work, we experimentally simulate and observe NH metals using single-photon interferometry (see Fig.~\ref{fig:setup} for an illustration of our setup), with a particular focus on the stability and topology of the ELs with respect to perturbations. Experimentally, this is achieved by implementing nonunitary time evolution for single photons that is governed by a corresponding NH Hamiltonian, and by performing interferometric measurements on the photons to extract the complex eigenenergies for each mode in reciprocal space. We simulate both NH metals featuring symmetry-protected ELs in two-dimensional (2D) systems and knotted ELs including a trefoil knot in three-dimensional (3D) systems, thus observing how EPs may form closed lines and even knots in momentum space.

More specifically, for symmetry-protected NH models in 2D, we show that the closed exceptional-line structures bound open Fermi volumes, and are robust against symmetry-preserving perturbations, while symmetry-breaking terms generically remove the ELs. This is in contrast to their 2D Hermitian analogs, where nodal points occur only at isolated momenta and do not give rise to bulk Fermi volumes. We further simulate 3D NH metals with both knotted and linked ELs that bound open Fermi surfaces representing the topological Seifert-surfaces of the corresponding knots~\cite{CSBB19}. There, observing the robustness of the band-structure topology with respect to generic perturbations, we confirm the topological stability of nodal knots in NH systems. Our results thus experimentally establish the topological variety and stability of nodal structures in NH metals.

{\it Theoretical framework.---}
We consider two-band NH metals, as described in reciprocal space by the NH Bloch Hamiltonian
\begin{align}
H(\textbf{k})=\textbf{d}_R(\textbf{k})\cdot\bm{\sigma}+i\textbf{d}_I(\textbf{k})\cdot\bm{\sigma},
\label{eq:Horg}
\end{align}
where $\bm{\sigma}$ are the standard Pauli matrices, $\textbf{k}$ is the lattice momentum for either a 2D or 3D lattice model, with the complex Bloch vector $\textbf{d}=\textbf{d}_R+i\textbf{d}_I$, where $\textbf{d}_R, \textbf{d}_I\in \mathbb{R}^3$.
With the eigenvalues of the NH Bloch Hamiltonian (\ref{eq:Horg}) given by $E_\pm=\pm\sqrt{d_R^2-d_I^2+2i\textbf{d}_R\cdot\textbf{d}_I}$, EPs  occur whenever eigenvalues coalesce ($E_+=E_-$) at nonvanishing $\textbf{d}$, i.e., for nontrivial solutions to the equations $d_R^2-d_I^2=0$ and $\textbf{d}_R\cdot\textbf{d}_I=0$~\cite{MA18}. Depending on the presence of NH symmetries as well as the spatial dimension of the system, the EPs can form closed ELs in the reciprocal space, with either symmetry-protected, or symmetry-independent topology. Specifically, ELs independent of symmetry occur in 3D, while suitable NH symmetries may reduce the codimension of EPs by one, thus stabilizing ELs in 2D systems. In both cases, ELs constitute boundaries for open Fermi volumes or surfaces, characterized by vanishing real parts of the energy gap, thus giving rise to symmetry-protected or intrinsic NH metals. Besides their topological stability, ELs in 3D may be topologically distinguished by forming different knots \cite{CSBB19}.

For the symmetry-protected NH metal, we consider a model on a 2D square lattice with unit lattice constant preserving the NH symmetry
\begin{equation}
H=qH^\dagger q^{-1}, q^\dagger q^{-1}=qq^\dagger=\one.
\label{eq:Q}
\end{equation}
By taking $q=\sigma_x$, the relation $\textbf{d}_R\cdot\textbf{d}_I=0$ is satisfied automatically~\cite{BCKB19}. The ELs are therefore defined and tunable as closed contours in 2D momentum space that satisfy the single constraint $d_R^2-d_I^2=0$. In particular, the parameter space is divided into regions with entirely real and entirely imaginary eigenspectrum, with ELs forming the boundary between the two.

By contrast, in three dimensions, the solutions of $\text{Re}(E^2)=0$ and $\text{Im}(E^2)=0$ each yield a
closed 2D surface in 3D momentum space, and these two hyperplanes generically intersect at topologically stable closed lines in the parameter space, thus giving rise to NH metals with ELs that are robust against symmetry-breaking perturbations. These ELs with intrinsic topology can form knots or links in reciprocal space, and are thus fundamentally distinct from symmetry-protected ELs in 2D.

{\it Experimental simulations.---}
We observe both symmetry-protected and intrinsic ELs by simulating the corresponding NH Bloch Hamiltonians $H(\mathbf{k})$ in reciprocal space, and by measuring the complex eigenenergies $E_\pm(\mathbf{k})$ using single-photon interferometry.
While an arbitrary NH dynamic is difficult to implement experimentally due to the difficulty of achieving gain in quantum systems~\cite{WKP+17}, especially with single photons, we circumvent this difficulty through a mapping
 \begin{equation}
H'(\mathbf{k})=H(\mathbf{k})+d_0\sigma_0,
\label{eq:Hprime}
\end{equation}
where $\sigma_0$ is the $2\times 2$ identity matrix, and $d_0=i\ln\sqrt{1/\Lambda}$, with $\Lambda=\max_{\mathbf{k}}|\lambda_{\mathbf{k}}|$ and $\lambda_{\mathbf{k}}$ the eigenvalue of $e^{-iH}e^{-iH^\dagger }$~\cite{Hal50,SMM18}. It follows that $H$ and $H'$ have the same eigenstates, while eigenenergies of $H'$ are related to those of $H$ through $E^\prime_\pm=E_\pm-i\ln\sqrt{\Lambda}$.

As a general framework, we encode the basis states into the orthogonal polarization states of a single photon, and initialize the polarization state in $\ket{\psi_\pm}$, the eigenstates of $H'(\mathbf{k})$ of a given $\mathbf{k}$ sector. We then send the photon through a $50:50$ beam splitter, after which the photon is in the state
\begin{equation}
\ket{\Psi_j}=\frac{1}{\sqrt{2}}\left(\ket{\psi_j}\ket{t}+\ket{\psi_j}\ket{r}\right),\quad (j=\pm)
\end{equation}
where $t$ and $r$ denote the transmitted and reflected modes of the single photon, respectively. The nonunitary time evolution governed by $e^{-iH'}$ is selectively enforced on the transmitted photon, leading to the state
\begin{align}
\ket{\Psi'_j}&=\left(e^{-iH^\prime}\otimes\ketbrad{t}+\one\otimes\ketbrad{r}\right)\ket{\Psi_j}\nonumber\\
&=\frac{1}{\sqrt{2}}\left(e^{-iE^\prime_j}\ket{\psi_j}\ket{t}+\ket{\psi_j}\ket{r}\right),
\end{align}
from which $E^\prime_j$ is extracted through an interferometric measurement~\cite{SM}.

\begin{figure*}
\includegraphics[width=0.65\textwidth]{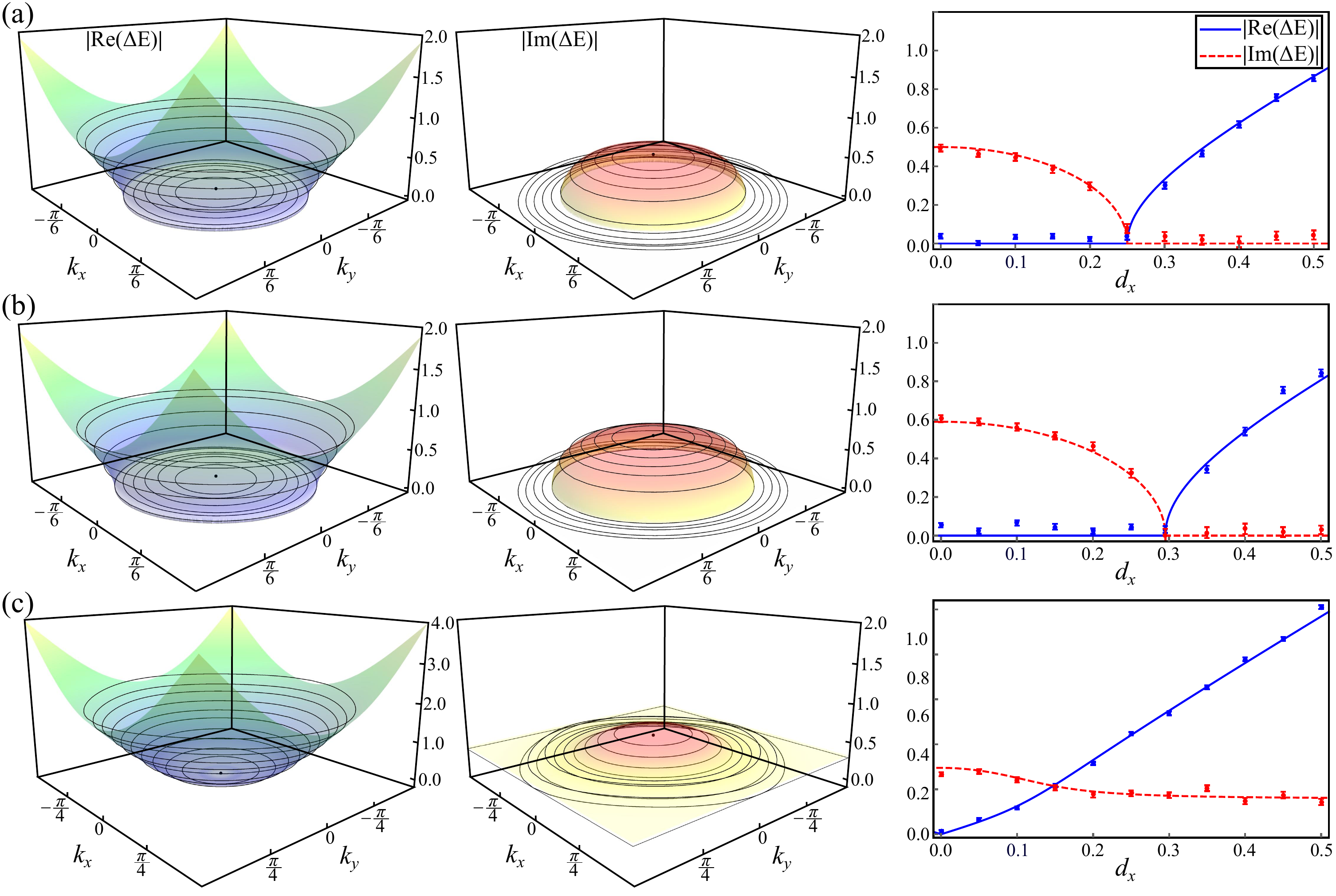}
   \caption{Observation of symmetry-protected ELs. (a) The real (left column) and imaginary (middle column) parts of the spectral gaps $\Delta E$ as a function of momentum for the symmetry-protected NH metal $H_1$. Experimental data are shown as black lines and theoretical results are shown as colored contour plots. Right column: the real and imaginary parts of the energy gap $\Delta E$ as a function of the parameter $d_x$. Theoretical predictions are represented by lines, and the experimental results by symbols. Error bars are obtained by assuming Poisson statistics in the photon-number fluctuations, indicating the statistical uncertainty.
   Effects due to the symmetry-preserving perturbation $i(\pi/20)\sigma_y$ and the symmetry-broken perturbation $i(\pi/20)\sigma_x$ are shown in (b) and (c), respectively.
   }
	\label{fig:2}
\end{figure*}

{\it Symmetry-protected ELs.---}
We first simulate the following 2D NH Hamiltonian with symmetry-protected ELs
\begin{equation}
\label{eq:H1}
H_1=(2-\cos k_x-\cos k_y)\sigma_x+\frac{i}{4}\sigma_z.
\end{equation}
Note that $H_1$ satisfies the symmetry defined in Eq.~(\ref{eq:Q}). We sample $11$ different Bloch vectors $d_x=2-\cos k_x-\cos k_y$, and measure the eigenenergies of both bands. Each sampled $d_x$ corresponds to a closed loop in the Brillouin zone that has the same eigenenergy.
In Fig.~\ref{fig:2}(a), we show the real and imaginary components of the energy gap $\Delta E=E_+-E_-$ in momentum space, which agree well with theoretical predictions. Particularly, an exceptional ring exists at $d_x=0.25$, by which the parameter space is divided into regions with either real or purely imaginary eigenenergies. It follows that the exceptional ring serves as a boundary for the open Fermi volume with $\text{Re}(\Delta E)=0$, i.e., an open set with the same dimensionality as the bulk system.
As a defining signature of symmetry protection, the topology of the exceptional ring is robust against symmetry-preserving perturbations. This is illustrated in Fig.~\ref{fig:2}(b), where the exceptional ring persists but is shifted in parameter space (now at $d_x=0.295$), upon the addition of a small perturbative term of the form $i(\pi/20)\sigma_y$. By contrast, when a symmetry-breaking perturbation of the form $i(\pi/20)\sigma_x$ is added [see Fig.~\ref{fig:2}(c)], EPs disappear, in accordance with theoretical predictions. Generalizing this strategy, we are able to experimentally simulate the general class of NH models with symmetry-protected ELs~\cite{SM}.

\begin{figure*}
\includegraphics[width=\textwidth]{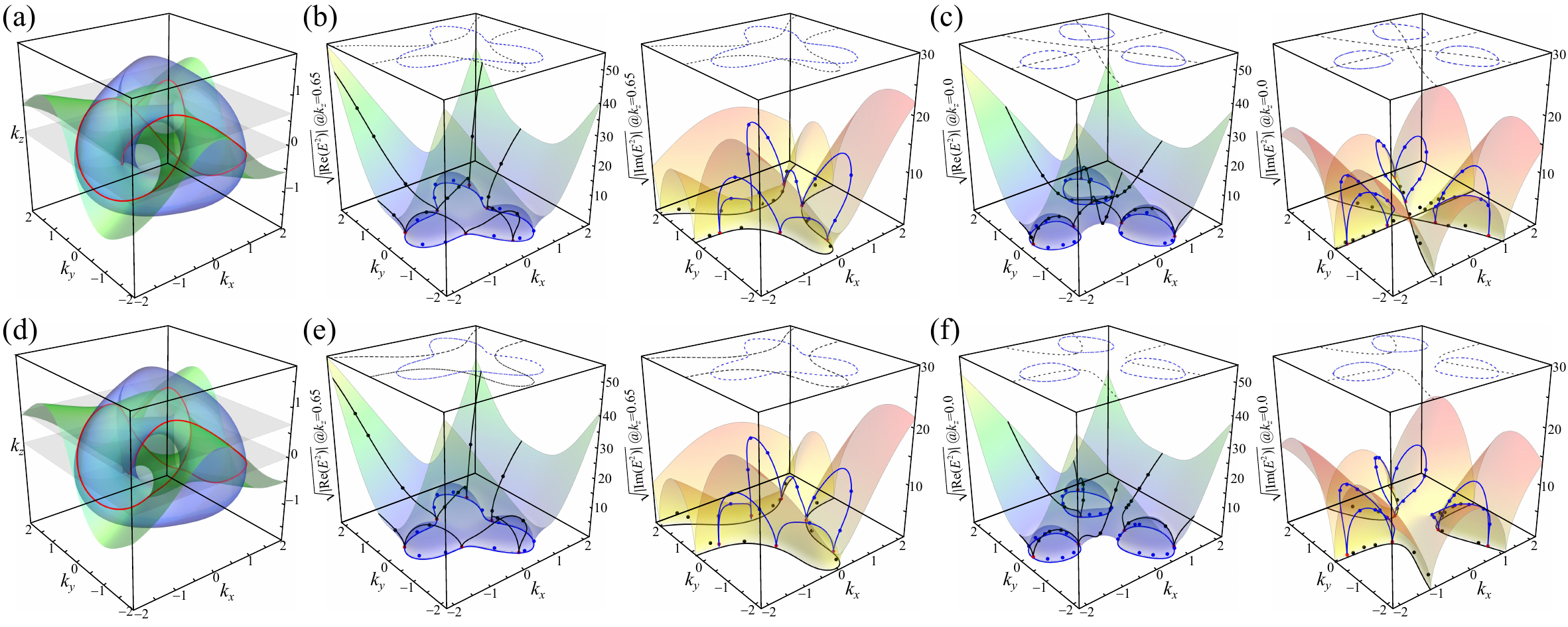}
   \caption{Observation of symmetry-independent knotted ELs. (a) Blue and green surfaces correspond to $\text{Re}(E^2)=0$ and $\text{Im}(E^2)=0$, respectively. The solid red curve is trefoil knotted EL, i.e., the intersection of two surfaces. Two gray planes correspond to the surfaces with $k_z=0.65$ and $0$, respectively, which are chosen in our experiment. $\sqrt{|\text{Re}(E^2)|}$ and $\sqrt{|\text{Im}(E^2)|}$ with fixed $k_z=0.65$ (b) and $k_z=0$ (c) as functions of $k_x$ and $k_y$. Solid black curves are the intersections between the surfaces $\text{Im}(E^2)=0$ and the gray planes. Blue solid curves are the intersections between the surfaces $\text{Re}(E^2)=0$ and the gray planes. EPs correspond to intersections of black and blue curves. Experimental data are shown as the black, blue, and red dots, and theoretical results are shown as the colored curves and the colored contour plots. (d)--(f) Effects of perturbations $\sum_{i=x,y,z}\delta_i\sigma_i$, where $\delta_i\in \left[0,0.4\right]$ are chosen randomly. In our experiment, we have $\delta_x=0.3179$, $\delta_y=0.3590$, and $\delta_z=0.2211$.}
   	\label{fig:5}
\end{figure*}

\begin{figure*}
\includegraphics[width=\textwidth]{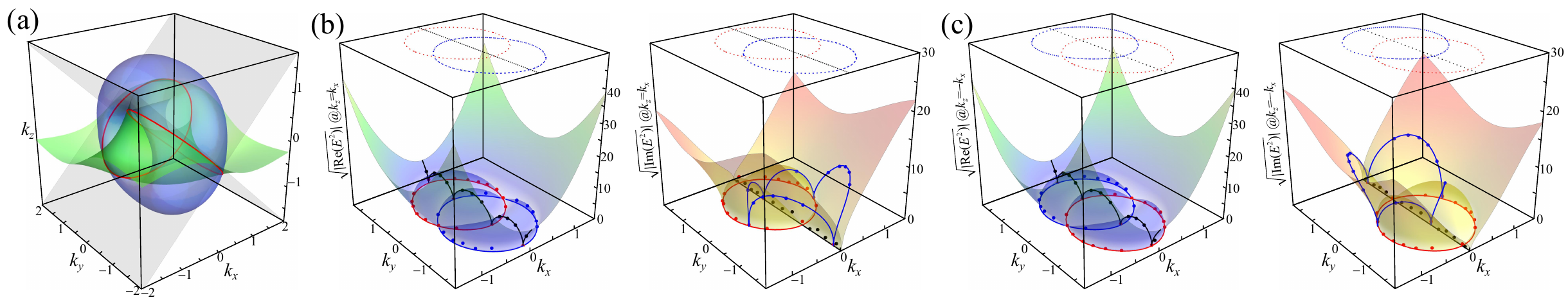}
   \caption{Observation of linked ELs. (a) Blue and green surfaces correspond to $\text{Re}(E^2)=0$ and $\text{Im}(E^2)=0$, respectively. The solid red curves correspond to the Hopf linked ELs, i.e., the intersection of two surfaces. The gray planes are the surfaces with $k_z=k_x,-k_x$, respectively, which are chosen in our experiment. $\sqrt{|\text{Re}(E^2)|}$ and $\sqrt{|\text{Im}(E^2)|}$ with fixed $k_z=k_x$ (b) and $k_z=-k_x$ (c) as functions of $k_x$ and $k_y$. Solid black curves are the intersections between the surfaces $\text{Im}(E^2)=0$ and the gray planes. Blue solid curves are the intersections between the surfaces $\text{Re}(E^2)=0$ and the gray planes. Red solid curves and the intersections of black and blue curves are ELs and EPs corresponding to the intersections between the Hopf link and the gray planes. The experimental data are shown as the black, blue, and red dots, and theoretical results are shown as the colored curves and the colored contour plots.
   }
	\label{fig:fig6}
\end{figure*}

{\it Symmetry-independent knotted ELs.---}
We now turn to NH models with symmetry-independent ELs. Following Ref.~\cite{CSBB19}, we construct models with intrinsic knotted or linked ELs in momentum space. This is achieved by taking
\begin{equation}
\label{eq:d}
\textbf{d}_R(\textbf{k})=\left[f_1(\textbf{k})-\epsilon,\epsilon,0\right], \textbf{d}_I(\textbf{k})=\left[0,f_2(\textbf{k}),\sqrt{2}\epsilon\right]
\end{equation}
in Eq.~(\ref{eq:Horg}), where $f_{1,2}(\textbf{k})$ are real and continuous scalar functions constrained by $f_1(\textbf{k})+if_2(\textbf{k})=Z_0^p+Z_1^q$, with $(Z_0,Z_1)\in\mathbb{C}^2$ and $|Z_0|^2+|Z_1|^2=1$. Here, $\epsilon$ is a constant with sufficiently large amplitude, which is fixed as $\epsilon=-20$ for our work. By construction, if $(p,q)$ are coprime  (both even) integers,
a NH Hamiltonian with Bloch vectors satisfying Eq.~(\ref{eq:d}) features ELs with $(p,q)$ knot (link) topology.

For our experiment, we adopt the construction
\begin{align}
\label{eq:construction}
&Z_0=\sin k_x+i\sin k_y,\\ &Z_1=2\sum_{\alpha=x,y,z} \cos k_\alpha-5+i\sin k_z\nonumber
\end{align}
to generate the functions $f_1$ and $f_2$ (such a construction is not unique though). First, we solve $f_1$ and $f_2$ using $(p,q)=(3,2)$, and simulate the corresponding NH Hamiltonian with trefoil-knotted ELs [see Fig.~\ref{fig:5}(a)]. In Figs.~\ref{fig:5}(b) and \ref{fig:5}(c), we show the measured $\sqrt{|\text{Re}(E^2)|}$ and $\sqrt{|\text{Im}(E^2)|}$ with fixed $k_z=0.65$  (b) and $k_z=0$ (c) as functions of $k_x$ and $k_y$, which agree well with the corresponding theoretical values and reveal the underlying knotted topology of the ELs. In this case, the knotted ELs serve as the boundary for the topologically nontrivial open Fermi-Seifert surface defined by $\text{Re}(E)=0$~\cite{CSBB19}.

To investigate the robustness of ELs, we introduce a general, symmetry-breaking perturbation $\sum_{i=x,y,z}\delta_i\sigma_i$, where $\delta_i\in \left[0,0.4\right]$ are chosen randomly. The experimental data shown in Figs.~\ref{fig:5}(d)--\ref{fig:5}(f) corresponds to the coefficients $\delta_x=0.3179$, $\delta_y=0.3590$, and $\delta_z=0.2211$. With the addition of perturbations, the knotted ELs still exist [see Figs.~\ref{fig:5}(d)--\ref{fig:5}(f)], while their shapes are slightly deformed compared to those without perturbations. These observations experimentally confirm and exemplify the robustness of knotted ELs to generic perturbations.

As a second case of NH model with intrinsic ELs, we adopt the same construction (\ref{eq:construction}), but with $(p,q)=(2,2)$. The resulting NH Hamiltonian features ELs with a link geometry [see Fig.~\ref{fig:fig6}(a)].
The experimentally measured $\sqrt{|\text{Re}(E^2)|}$ and $\sqrt{|\text{Im}(E^2)|}$ with $k_z=\pm k_x$ are shown in Figs.~\ref{fig:fig6}(b) and \ref{fig:fig6}(c). Similar to the knotted ELs, the linked topology here is also found to be robust to generic perturbations.

{\it Conclusion.---}
By simulating and observing two different classes of NH metals, our present experimental study corroborates the topological robustness of ELs that occur in a rich variety of nodal NH band structures. While our findings hopefully inspire the investigation of exotic NH metals also in other physical platforms, we note that our present experimental scheme may readily be extended to multiband models and systems with other symmetries, thus offering a versatile toolbox for the systematic experimental study of nodal phases in both Hermitian and non-Hermitian settings. Furthermore, our configuration enables
the investigation of dynamical properties of NH metals, where the presence of open Fermi volumes or open Fermi surfaces may give rise to so far unexplored phenomena.

\begin{acknowledgments}
We acknowledge fruitful discussions with Emil J. Bergholtz. This work has been supported by the Natural Science Foundation of China (Grants No. 12025401, No. 11974331, and No. U1930402). L. X. acknowledges support from the project funded by China Postdoctoral Science Foundation (Grant No. 2020M680006). W. Y. acknowledges support from the National Key Research and Development Program of China (Grants No.
2016YFA0301700 and No. 2017YFA0304100). J. C. B. acknowledges financial support from the German Research Foundation (DFG) through the Collaborative Research Centre SFB 1143 (project ID 247310070) and the Cluster of Excellence ct.qmat (EXC 2147, project ID 390858490).
\end{acknowledgments}

\bibliography{NH_Metals.bib}

\clearpage
\begin{widetext}
\appendix

\renewcommand{\thesection}{\Alph{section}}
\renewcommand{\thefigure}{S\arabic{figure}}
\renewcommand{\thetable}{S\Roman{table}}
\setcounter{figure}{0}
\renewcommand{\theequation}{S\arabic{equation}}
\setcounter{equation}{0}

\section{SUPPLEMENTAL MATERIAL FOR ``SIMULATING EXCEPTIONAL NON-HERMITIAN METALS
WITH SINGLE-PHOTON INTERFEROMETRY"}

\subsection{Experimental detection}

A qubit is encoded in the polarization of a single photon $\{\ket{H}=(1,0)^\text{T},\ket{V}=(0,1)^\text{T}\}$, generated by the standard spontaneous parametric down-conversion, and projected into the eigenstate $\ket{\psi_\pm}$ of $H'(k)$ with a polarizing beam splitter (PBS), a half-wave plate (HWP) and a quarter-wave plate (QWP). 

Following the $50:50$ beam splitter (BS), the photon state is given by
\begin{equation}
\ket{\Psi_j}=\frac{1}{\sqrt{2}}\ket{\psi_j}\left(\ket{t}+\ket{r}\right), j=\pm,
\end{equation}
where $t$ and $r$ denote the transmission and reflection modes of the single photons, respectively. To measure the complex phase shift $E^\prime_\pm$ under $U'=e^{-iH'}$,
we perform the nonunitary operation $U'$ on the polarization state of the photons in the transmission mode and keep that in the reflected mode unchanged, i.e., $U'\otimes \ket{t}\bra{t}+\one\otimes\ket{r}\bra{r}$, where $\one$ is the identity operator. The nonunitary operation $U'$ can be realized by an interferometric network involving beam displacers (BDs) and wave plates.

As illustrated in Fig.~1 of the main text, in our experiment we decompose $U'$ according to $U'=R_2L(\theta_V)R_1$~\cite{LWZ19}, where an arbitrary rotation of $R_j$ ($j=1,2$) is realized by two QWPs and a HWP. Two BDs and two HWPs with setting angles $45^\circ$ and $\theta_V$ are used to realize the polarization-dependent loss operator $L=\ket{V}\bra{H}+l\ket{H}\bra{V}$, where $\theta_V=\frac{\arcsin l}{2}$ is determined by the loss parameter of $l$.

The state then evolves according to
\begin{equation}
\ket{\Psi'_j}=\frac{1}{\sqrt{2}}\left(e^{-iE^\prime_j}\ket{\psi_j}\ket{t}+\ket{\psi_j}\ket{r}\right).
\end{equation}
We then perform interferometric measurements~\cite{WQX19}: a HWP at $45^\circ$ is applied on the polarizations of the photons in the transmitted mode. After passing through a polarizing beam splitter (PBS), the state evolves into
\begin{align}
\ket{\Psi''_j}=&\frac{\alpha_j}{\sqrt{2}}\left(\ket{H}+e^{-iE^\prime_j}\ket{V}\right)\ket{t'}+\frac{\beta_j}{\sqrt{2}}\left(e^{-iE^\prime_j}\ket{H}+\ket{V}\right)\ket{r'},
\end{align}
under the assumption that $\ket{\psi_j}=\alpha_j\ket{H}+\beta_j\ket{V}$. Here $t'$ ($r'$) denotes the transmitted (reflected) mode following the second PBS in the path of the $r$-mode photons.
We perform projective measurements with the bases $\{\ket{H},\ket{V},\ket{+}=(\ket{H}+\ket{V})/\sqrt{2},\ket{R}=(\ket{H}-i\ket{V})/\sqrt{2}\}$ on the polarizations of the photons in either of the transmitted ($t'$) or reflected ($r'$) mode. The outputs are recorded in coincidence with trigger photons. The coincidences are denoted as $\{N_H,N_V,N_+,N_R\}$, which satisfy the following relations
\begin{align}
& N_H=\mathcal{N} \frac{|\alpha_j|^2}{2}, N_V=\mathcal{N} \frac{|\alpha_j|^2}{2}e^{2\text{Im}(E^\prime_j)},\nonumber\\
& N_+=N_H\left(\frac{1}{2}e^{i E^\prime_j}+\frac{1}{2}e^{-i E^\prime_j}+\frac{1}{2}+\frac{1}{2}e^{2\text{Im}(E^\prime_j)}\right),\nonumber\\
& N_R=N_H\left(\frac{i}{2}e^{-i E^\prime_j}-\frac{i}{2}e^{i E^\prime_j}+\frac{1}{2}+\frac{1}{2}e^{2\text{Im}(E^\prime_j)}\right),
\end{align}
where $\mathcal{N}$ corresponds to the number of photons in $r$ mode. We then obtain $E^\prime_j$ through
\begin{equation}
E^\prime_j=i\text{ln}\left(\frac{2N_+-N_H-N_V}{2N_H}+i\frac{N_H+N_V-2N_R}{2N_H}\right)+2n\pi,
\end{equation}
when the polarization of photons in the $t'$ mode is measured. Similarly, when the polarization of the photons in the $r'$ mode is measured, we have
\begin{equation}
E^\prime_j=i\text{ln}\left(\frac{2N_+-N_H-N_V}{2N_V}+i\frac{2N_R-N_H-N_V}{2N_V}\right)+2n\pi,
\end{equation}
where $n$ is an integer. From the relation $E^\prime_j=E_j-i\ln\sqrt{\Lambda}$, one can also obtain the eigenenergies of $H(k)$.

Here we give a concrete example of our experimental process. For a 2D non-Hermitian Hamiltonian with symmetry-protected exceptional lines $H_1$ in Eq.~(6) of the main text, with Bloch vector $d_x=2-\cos k_x-\cos k_y=0.2$. The corresponding nonunitary evolution with unit time is then $U=\begin{pmatrix}
                                      1.262 & -0.201i \\
                                      -0.201i & 0.760
                                    \end{pmatrix}$.
Instead, in our experiment we realize the passive nonunitary operation $U'=\begin{pmatrix}
                                      0.985 & -0.157i \\
                                      -0.157i & 0.593
                                    \end{pmatrix}$ governed by
the effective Hamiltonian $H^\prime_1=H_1+i\ln\sqrt{1/\Lambda}\sigma_0=\begin{pmatrix}
                                      0.002i & 0.200 \\
                                      0.200 & -0.498i
                                    \end{pmatrix}$ with $\Lambda=1.643$.
The heralded single photons are prepared in the state $\ket{\psi_-}$ via a HWP at $-31.7^\circ$ and a QWP at $0^\circ$ as the operators of HWPs and QWPs are $\text{HWP}(\theta)=\begin{pmatrix}
                                      \cos2\theta & \sin2\theta \\
                                      \sin2\theta & -\cos2\theta
                                    \end{pmatrix}$ and $\text{QWP}(\theta)=\begin{pmatrix}
                                      \cos\theta^2+i\sin\theta^2 & (1-i)\sin\theta\cos\theta \\
                                      (1-i)\sin\theta\cos\theta & \sin\theta^2+i\cos\theta^2
                                    \end{pmatrix}$, respectively.
The passive nonunitary operation is decomposed as $U'=R_2LR_1$, where $R_1=\begin{pmatrix}
                                      -0.995 & 0.098i \\
                                      -0.098 & -0.995i
                                    \end{pmatrix}$, $L=\begin{pmatrix}
                                      0 & 0.609 \\
                                      1 & 0
                                    \end{pmatrix}$, and $R_2=\begin{pmatrix}
                                      0.098 & -0.995 \\
                                      0.995i & 0.098i
                                    \end{pmatrix}$.
To implement it, all the QWPs are set as $0^\circ$ and the HWPs are set at $-2.8^\circ$, $\theta_V=18.7^\circ$ and $-47.8^\circ$ (from left to right, Fig.~1 in the main text), respectively.

To verify the accuracy of the nonunitary operation in our experiment, we adopt a quantity that characterizes the distance between the implemented and the target operations. In general,
an operation in a quantum system can be expressed in the operator-sum representation as a positive-definite process matrix $\chi$ which
relates the output state $\rho_{out}$ with the input state $\rho_{in}$, through
\begin{equation}
\rho_{out}=\sum_{m,n=1}^{4}\chi_{mn}E_m\rho_{in}E_n^{\dag}:==\mathcal{E}(\rho_{in}),
\end{equation}
where $\chi_{mn}$ are the elements of the process matrix $\chi$, $\{E_m\}$ corresponds to the standard Pauli matrix in general dimensions,
and $\mathcal{E}$ stands for the quantum process.
We then define a distance $d$, with
\begin{equation}
d=\frac{1}{2}\text{Tr}\left[(\frac{\chi_{\text{e}}}{\text{Tr}(\chi_{\text{e}})}-\frac{\chi_{\text{t}}}{\text{Tr}(\chi_{\text{t}})})(\frac{\chi_{\text{e}}}{\text{Tr}(\chi_{\text{e}})}-\frac{\chi_{\text{t}}}{\text{Tr}(\chi_{\text{t}})})^\dagger\right],
\end{equation}
where $\chi_{\text{e}}$ and $\chi_{\text{t}}$ are, respectively, the measured process matrix via process tomography~\cite{WWZ18} and its theoretical predictions. The distance $d$ varies between $0$ and $1$, which can be used to quantify the match between the measured process matrix and its theoretical prediction~\cite{GLN05}, with $0$ indicating a perfect match and $1$ indicating a complete mismatch. For our experiments, we obtain $d=0.0076\pm0.0005$, indicating an excellent match.

Alternatively, we may define the distance based on the Jamiolkowski isomorphism $\rho_{\mathcal{E}}$ of a quantum process $\mathcal{E}$~\cite{Jmatrix72}
\begin{equation}
\rho_{\mathcal{E}}=(\one\otimes\mathcal{E})(\ketbrad{\Psi}),
\end{equation}
where $\ket{\Psi}=\sum_{i=0}^{d-1}\ket{i}\ket{i}/\sqrt{d}$ is the maximally entangled
state of the system and a $d$-dimensional ancilla ($d$ is the dimension of the system), and $\{\ket{i}\}$ is the computational basis. The distance between the measured and ideal processes can be quantified using the trace distance
\begin{equation}
D=\frac{1}{2}\text{Tr}\left[\sqrt{(\rho_{\mathcal{E_\text{e}}}-\rho_\mathcal{E_\text{t}})^\dagger(\rho_{\mathcal{E_\text{e}}}-\rho_\mathcal{E_\text{t}})}\right],
\end{equation}
where $\rho_{\mathcal{E_\text{e}}}$ and $\rho_{\mathcal{E_\text{t}}}$ are, respectively, the state obtained via the process tomography and its theoretical prediction. It has been shown that $D$ serves as the upper bound on the distance between the measured and ideal joint probability distributions during a sampling computation~\cite{GLN05}.
For our experiments, we obtain $D=0.0695\pm0.0020$, again indicating an excellent match.

\subsection{Alternative model with symmetry-protected exceptional lines}

\begin{figure*}
\includegraphics[width=1\textwidth]{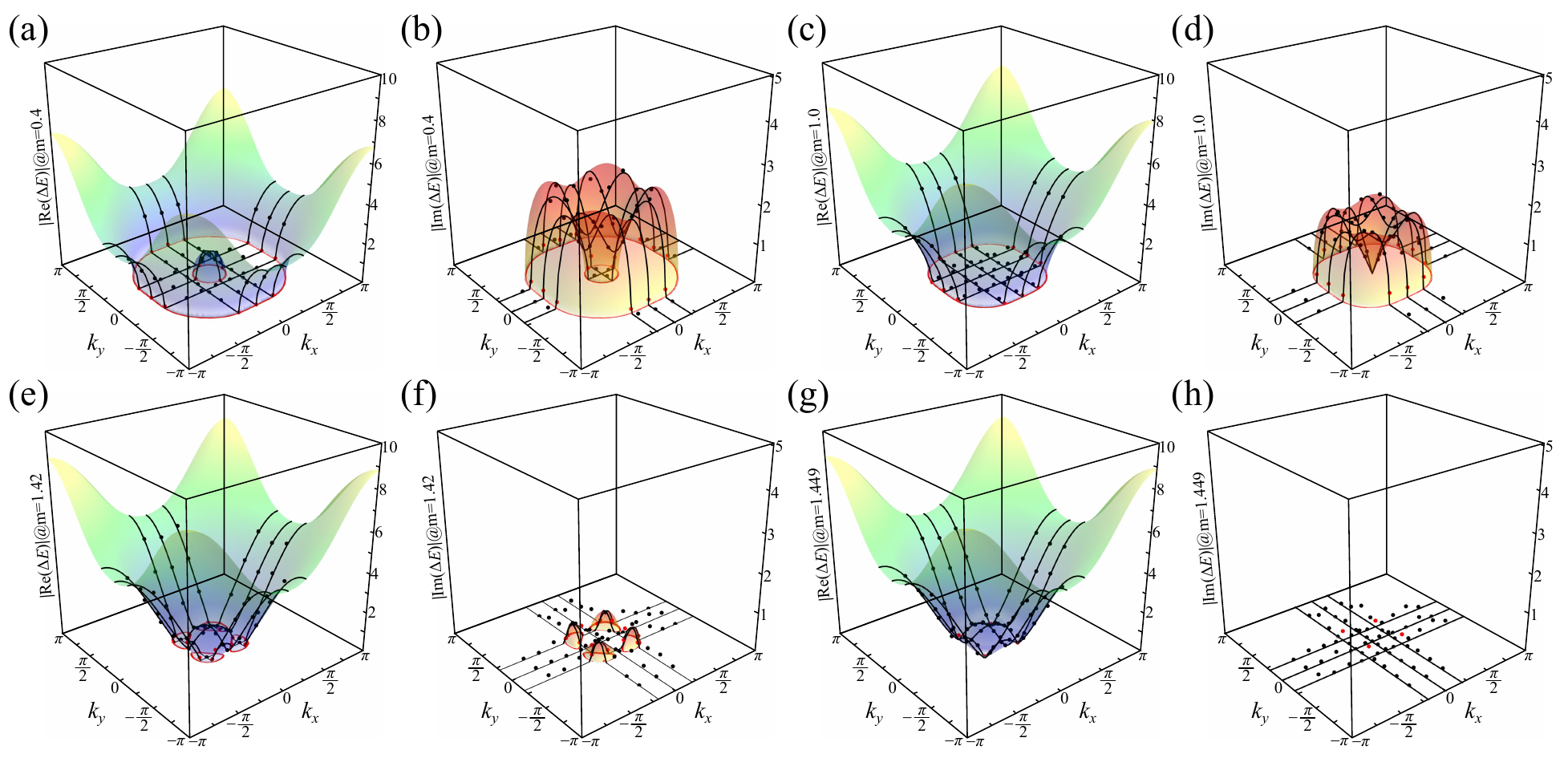}
   \caption{The real and imaginary parts of the spectral gaps $\Delta E$ of the Hamiltonian $H_2$ in Eq.~(\ref{eq:H2}) with different $m$. (a-b) $m=0.4$. (c-d) $m=1$. (e-f) $m=1.42$. (g-h) $m=\sqrt{6}-1$. Red solid curves are exceptional lines and solid black curves correspond to the theoretical results with different fixed $k_x$ ($k_y$) sectors.
   Experimental data are shown as the red and black dots.}
	\label{fig:3}
\end{figure*}

\begin{figure*}[htb]
\includegraphics[width=0.75\textwidth]{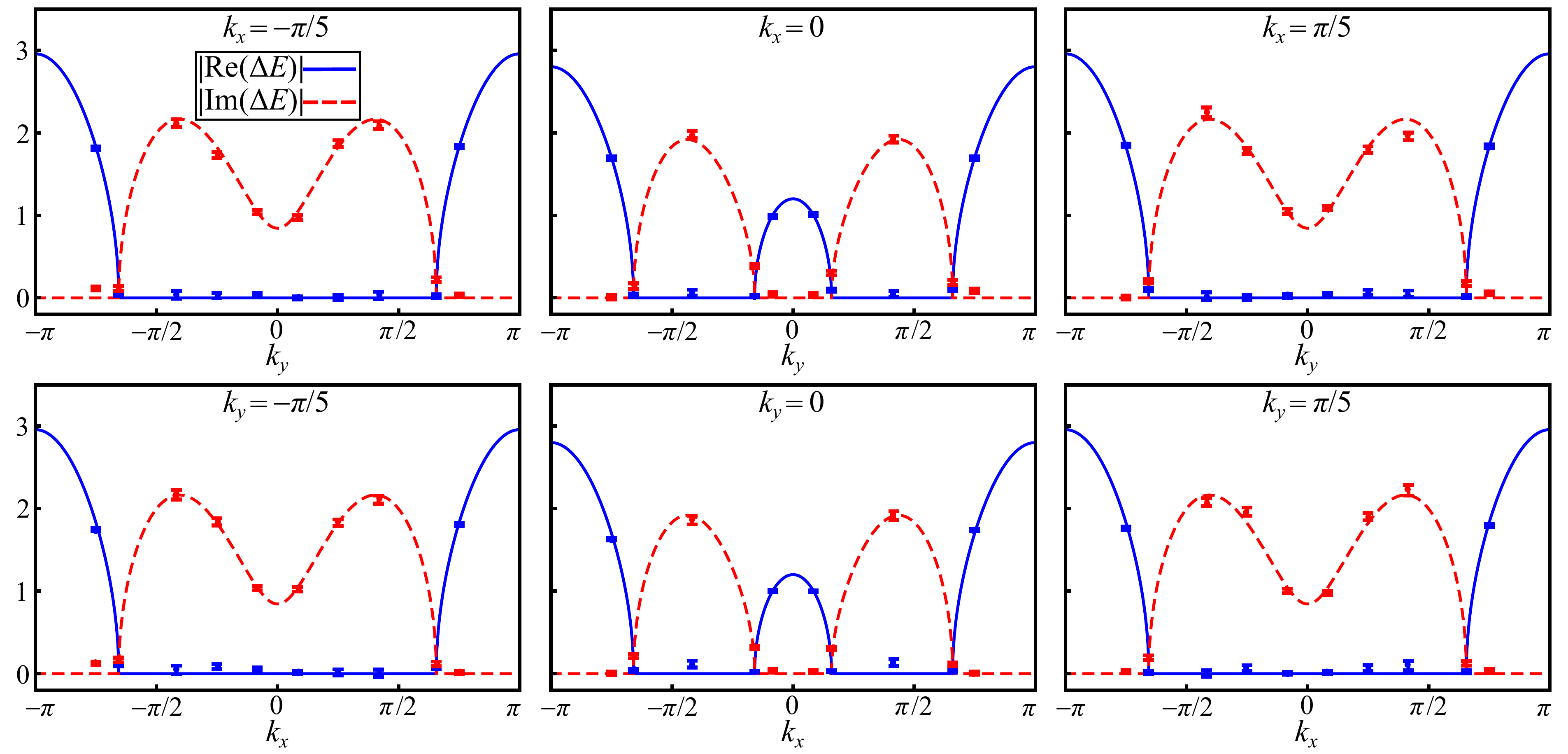}
   \caption{The real and imaginary parts of the energy gap $\Delta E$ of a generic symmetry-protected non-Hermitian nodal phase of the model $H_2$ with fixed $m=0.4$ and fixed $k_x (k_y)=-\pi/5,0,\pi/5$ as a function of the parameter $k_y$ ($k_x$). Theoretical predictions are represented by solid blue and dashed red curves, and the experimental results by symbols. Error bars indicate the statistical uncertainty, obtained by assuming Poissonian statistics in the photon-number fluctuations.}
	\label{fig:S1}
\end{figure*}

\begin{figure*}[htb]
\includegraphics[width=0.75\textwidth]{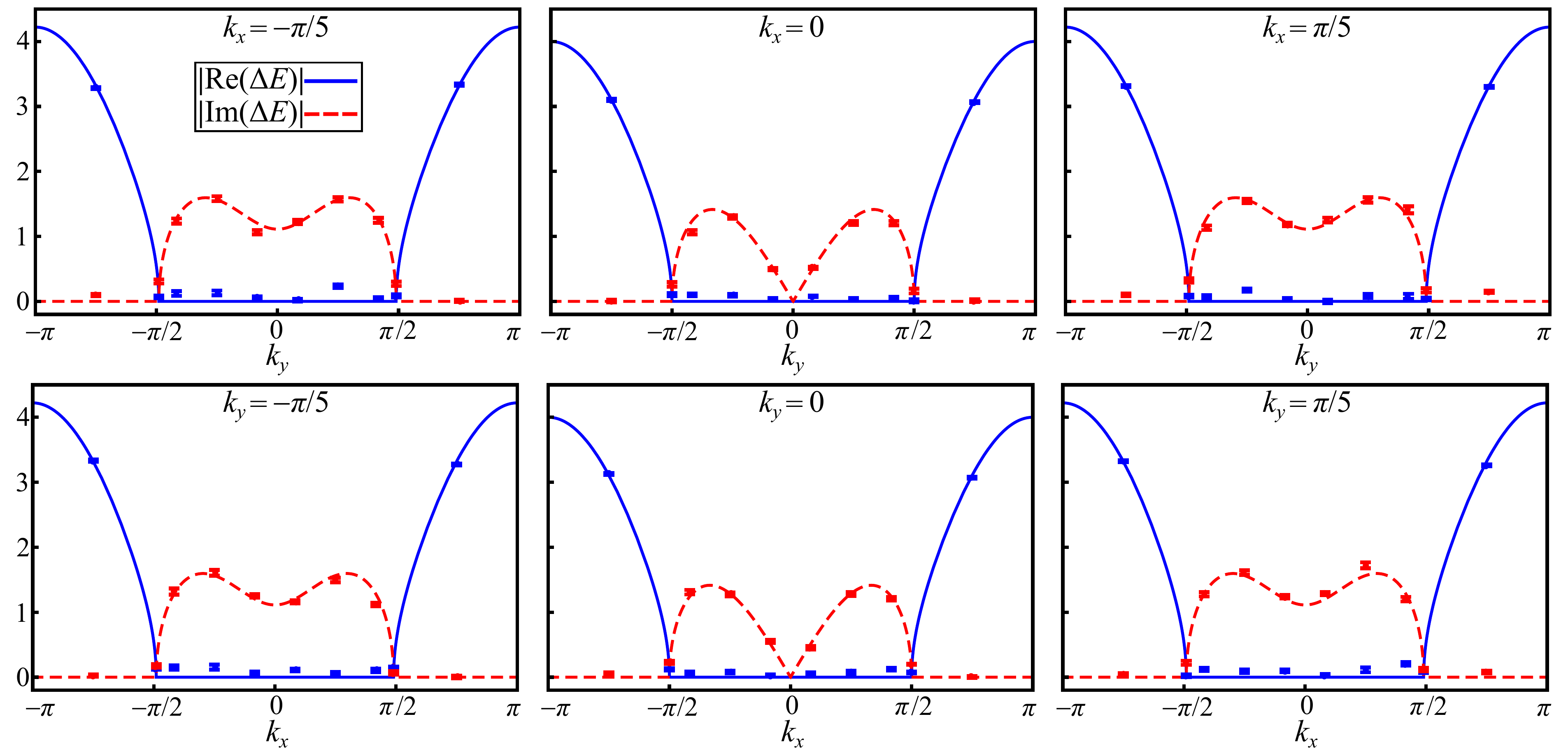}
   \caption{The real and imaginary parts of the energy gap $\Delta E$ of a generic symmetry-protected non-Hermitian nodal phase of the model $H_2$ with fixed $m=1$ and fixed $k_x (k_y)=-\pi/5,0,\pi/5$ as a function of the parameter $k_y$ ($k_x$). Theoretical predictions are represented by solid blue and dashed red curves, and the experimental results by symbols. Error bars indicate the statistical uncertainty, obtained by assuming Poissonian statistics in the photon-number fluctuations.}
	\label{fig:S2}
\end{figure*}

\begin{figure*}[htb]
\includegraphics[width=0.75\textwidth]{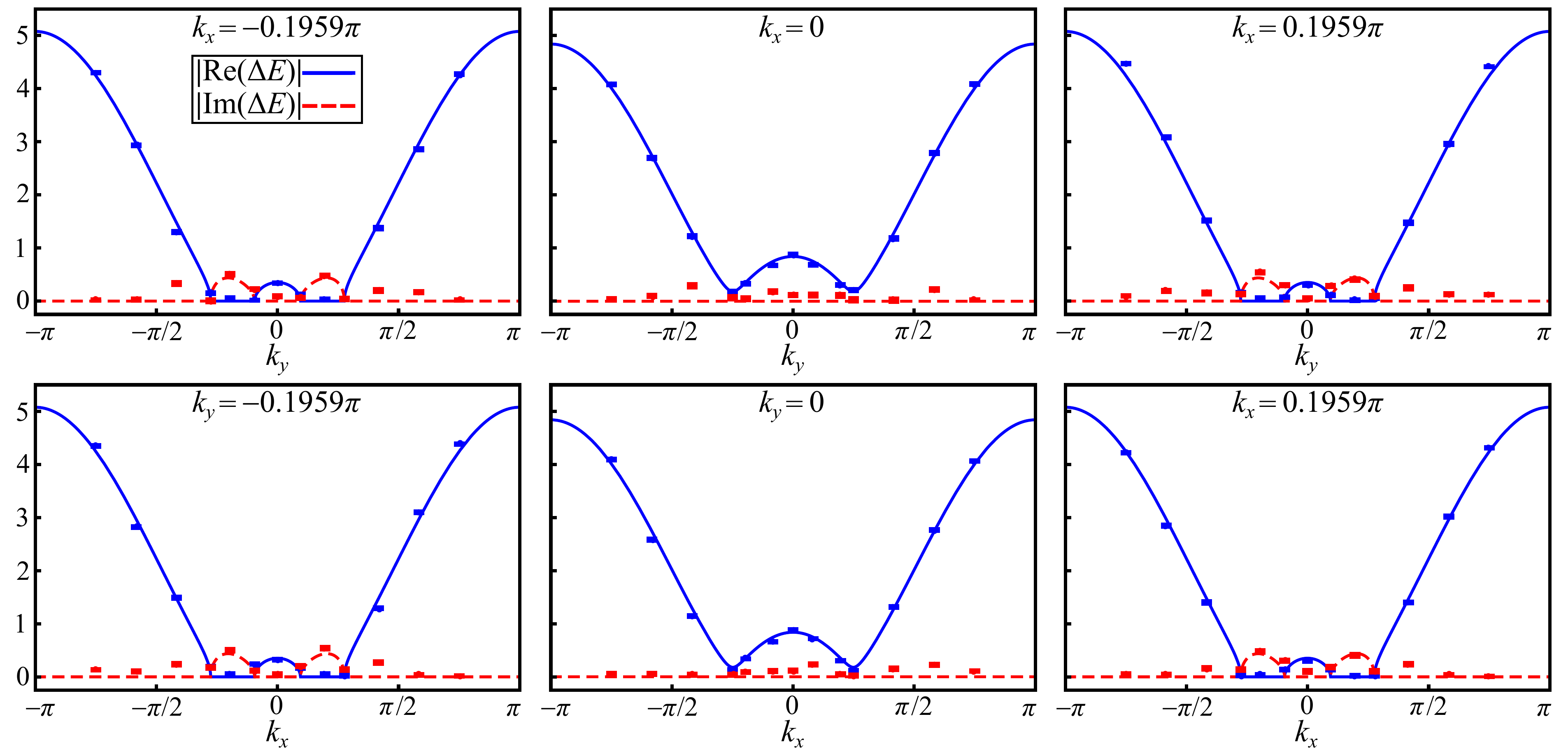}
   \caption{The real and imaginary parts of the energy gap $\Delta E$ of a generic symmetry-protected non-Hermitian nodal phase of the model $H_2$ with fixed $m=1.42$ and fixed $k_x (k_y)=-0.1959\pi,0,0.1959\pi$ as a function of the parameter $k_y$ ($k_x$). Theoretical predictions are represented by solid blue and dashed red curves, and the experimental results by symbols. Error bars indicate the statistical uncertainty, obtained by assuming Poissonian statistics in the photon-number fluctuations.}
	\label{fig:S3}
\end{figure*}

\begin{figure*}[htb]
\includegraphics[width=0.75\textwidth]{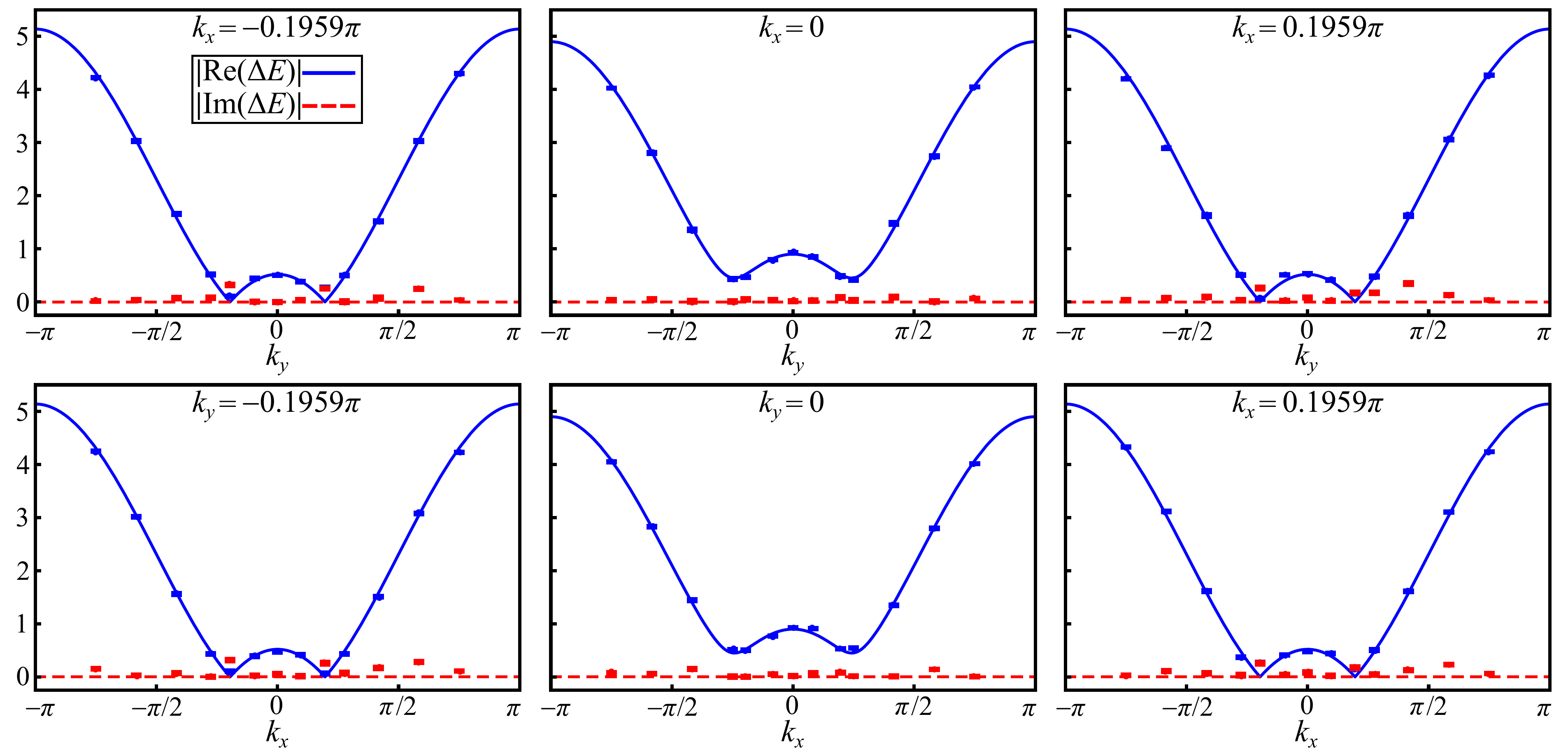}
   \caption{The real and imaginary parts of the energy gap $\Delta E$ of a generic symmetry-protected non-Hermitian nodal phase of the model $H_2$ with fixed $m=\sqrt{6}-1$ and fixed $k_x (k_y)=-0.1959\pi,0,0.1959\pi$ as a function of the parameter $k_y$ ($k_x$). Theoretical predictions are represented by solid blue and dashed red curves, and the experimental results by symbols. Error bars indicate the statistical uncertainty, obtained by assuming Poissonian statistics in the photon-number fluctuations.}
	\label{fig:S4}
\end{figure*}

As a complement to $H_1$ in the main text, we consider an alternative 2D model with symmetry-protected exceptional lines
\begin{equation}
H_2=\left(m+1-\cos k_x-\cos k_y\right)\sigma_x+i\sin k_x\sigma_y+i\sin k_y\sigma_z,
\label{eq:H2}
\end{equation}
where $m\in \mathbb{R}$. Note that $H_2$ also satisfies the symmetry in Eq.~(2) (in the main text) with $q=\sigma_x$. In Figs.~\ref{fig:3}(a) and (b), we choose $m=0.4$, satisfying the condition $|m+1|<1+\sqrt{2}$. We then sample $60$ $\mathbf{k}$ sectors in the Brillouin zone (dots in Fig.~\ref{fig:3}), with
$k_x(k_y)=\pm \pi/5$, $k_y(k_x)=\{\pm3\pi/4,$ $\pm0.6556\pi,$ $\pm5\pi/12,$ $\pm\pi/4,$ $\pm\pi/12\}$, and $k_x(k_y)=0$, $k_y(k_x)=\{\pm 3\pi/4,$ $\pm0.6587\pi,$ $\pm5\pi/12,$ $\pm0.1587\pi,$ $\pm\pi/12\}$. By observing the behavior of $\text{Re}(\Delta E)$ and $\text{Im}(\Delta E)$ (see Fig.~\ref{fig:S1}), we are able to infer the locations of exceptional points, which form a pair of closed loops in the momentum space, consistent with theoretical predictions. The closed exceptional lines form the boundary for open 2D Fermi surfaces.

Similarly, in Figs.~\ref{fig:3}(c) and (d) [see also Fig.~\ref{fig:S2}], we choose $m=1$, satisfying $|m+1|=2$. We then sample the Brillouin zone: $k_x(k_y)=\pm\pi/5$, $k_y(k_x)=\{\pm3\pi/4,$ $\pm5\pi/12,$ $\pm0.49\pi,$ $\pm\pi/4,$ $\pm\pi/12\}$, and $k_x(k_y)=0$, $k_y(k_x)=\{\pm3\pi/4,$ $\pm5\pi/12,$ $\pm\pi/2,$ $\pm\pi/4,$ $\pm\pi/12\}$. The inner exceptional line in this case collapses into a single exceptional point. Further increase in $m$ [see Figs.~\ref{fig:3}(e) and (f); see also Fig.~\ref{fig:S3}] leads to the splitting of the outer exceptional line into four Fermi pockets. Finally, for sufficiently large $m$ [see Figs.~\ref{fig:3}(g) and (h); see also Fig.~\ref{fig:S4}], all four exceptional lines collapse to four exceptional points.
We have checked that the existence of exceptional lines or exceptional points are also robust against symmetry-preserving perturbations.

\subsection{Further information of intrinsic exceptional lines}

\begin{figure*}[htb]
\includegraphics[width=0.9\textwidth]{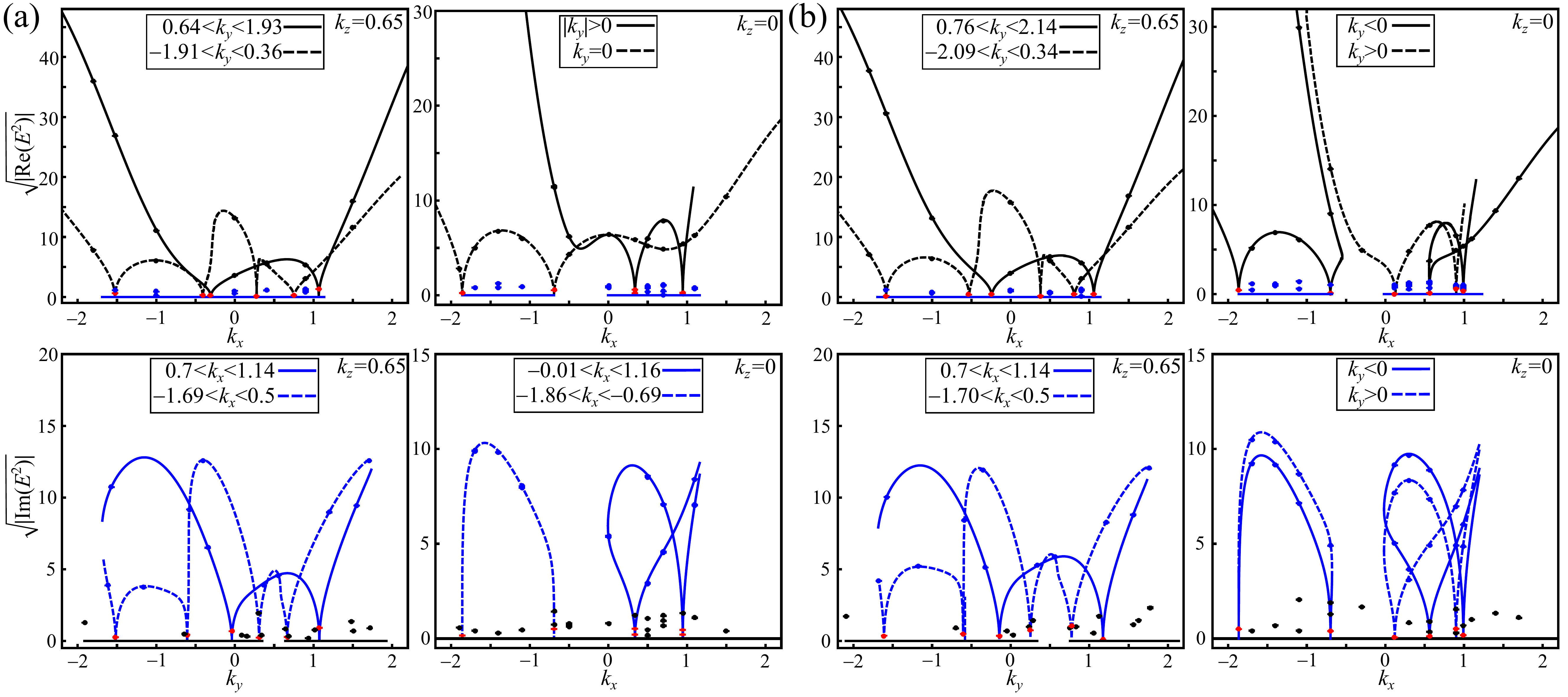}
   \caption{The cross-sectional view of $\sqrt{|\text{Re}(E^2)|}$ and $\sqrt{|\text{Im}(E^2)|}$ of the knotted nodal band structure obtained from the model defined by Eq.~(7) (a) and Eq.~(7) with symmetry-breaking perturbations of $\sum_{i=x,y,z}\delta_i\sigma_i$ (b) with $(p,q)=(3,2)$, $\epsilon=-20$ and $k_z=0.65,0$. Theoretical predictions are represented by curves, and the experimental results by symbols. Error bars indicate the statistical uncertainty, obtained by assuming Poissonian statistics in the photon-number fluctuations.}
	\label{fig:S5}
\end{figure*}

\begin{figure*}[htb]
\includegraphics[width=0.9\textwidth]{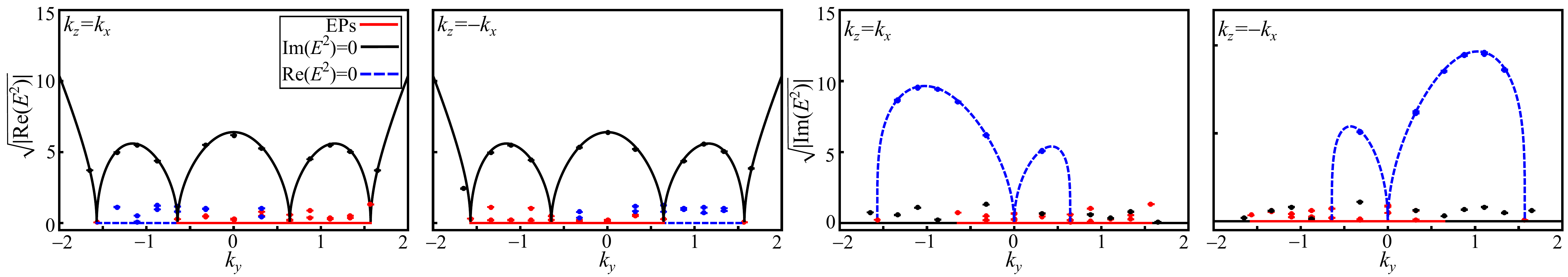}
   \caption{The cross-sectional view of $\sqrt{|\text{Re}(E^2)|}$ and $\sqrt{|\text{Im}(E^2)|}$ of the linked nodal band structure obtained from the model defined by Eq.~(7) (in the main text) with $(p,q)=(2,2)$, $\epsilon=-20$ and $k_z=\pm k_x$. Theoretical predictions are represented by curves, and the experimental results by symbols. Error bars indicate the statistical uncertainty, obtained by assuming Poissonian statistics in the photon-number fluctuations.}
	\label{fig:S6}
\end{figure*}

In Figs.~\ref{fig:S5} and~\ref{fig:S6}, we provide cross-sectional view for experimental data points in Figs.~3 and 4 of the main text. Crucially, whereas eigenenergie can be significantly affected by perturbations, the topology of exceptional lines are unchanged.

\begin{figure*}
\includegraphics[width=\textwidth]{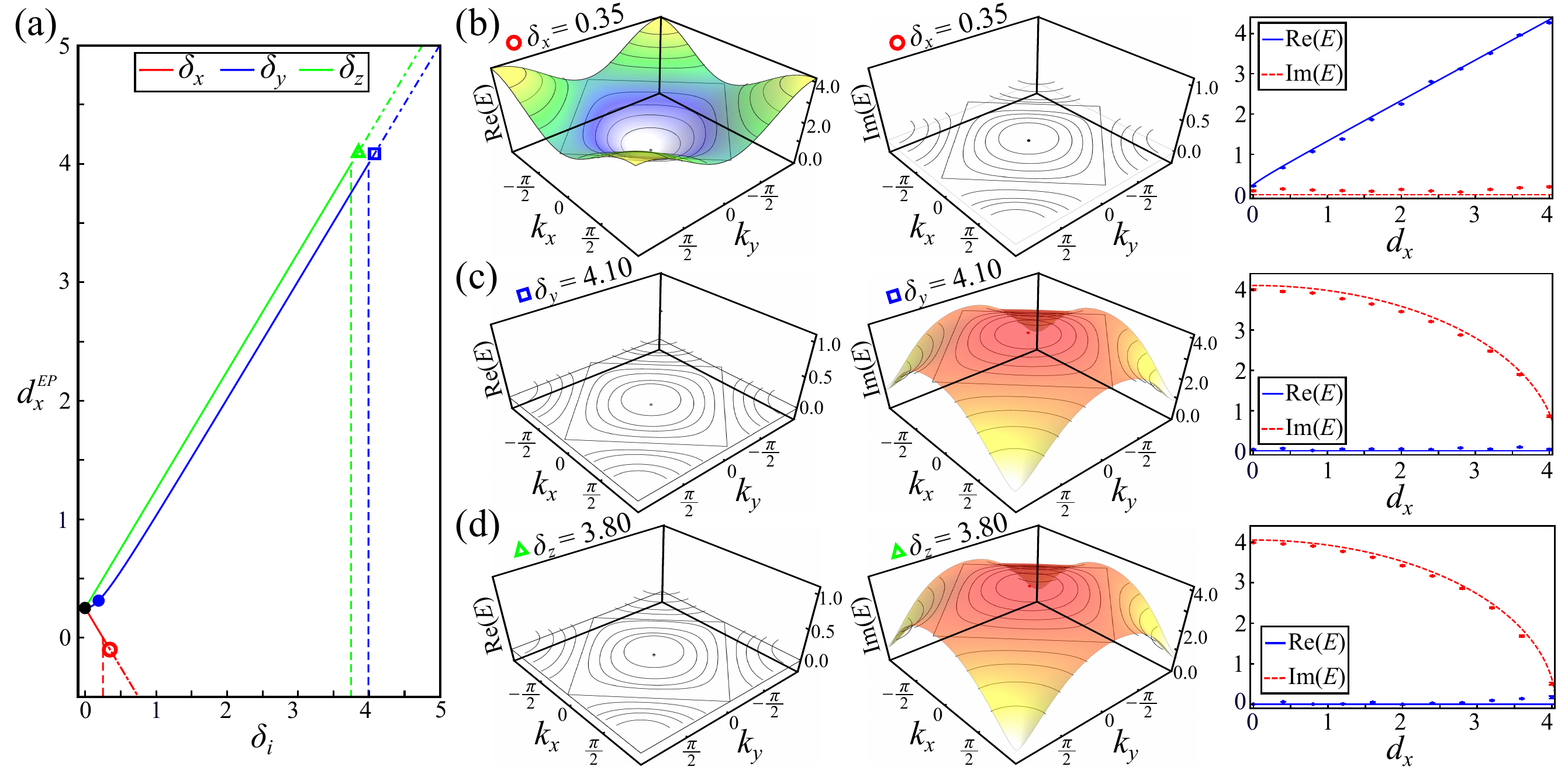}
   \caption{Robustness of symmetry-protected exceptional lines versus perturbation. (a) The strength of $d_x$ of the exceptional line versus the perturbation of $i\delta_x\sigma_x$  (red solid line), $\delta_y\sigma_y$ (blue solid line), and $\delta_z\sigma_z$ (green solid line), respectively. The critical points, where the knotted exceptional lines disappear, are $\{\delta_x=0.25, \delta_y=3.99, \delta_z=3.75\}$ as indicated by vertical dashed lines. The black and blue filled dots correspond to the parameters without perturbation and with symmetry-preserving perturbation $i\frac{\pi}{20}\sigma_y$, where the exceptional lines still exist. The empty circle, square and triangle located at $\delta_x=0.35$, $\delta_y=4.10$ and $\delta_z=3.80$ indicate the perturbation parameters for (b), (c) and (d), respectively. (b)-(d) The real (left column) and imaginary (middle column) parts of energy $E$ as functions of momentum for $H_1$ with different symmetry-preserving perturbations. Experimental data are shown as black lines, and theoretical results are shown as colored contour plots. Right column: The real and imaginary parts of the energy $E$ as functions of the parameter $d_x$. Theoretical predictions are represented by lines, and the experimental results by symbols. Error bars are obtained by assuming Poisson statistics in the photon-number fluctuations, indicating the statistical uncertainty.
}
	\label{fig:S8}
\end{figure*}

\begin{figure*}
\includegraphics[width=\textwidth]{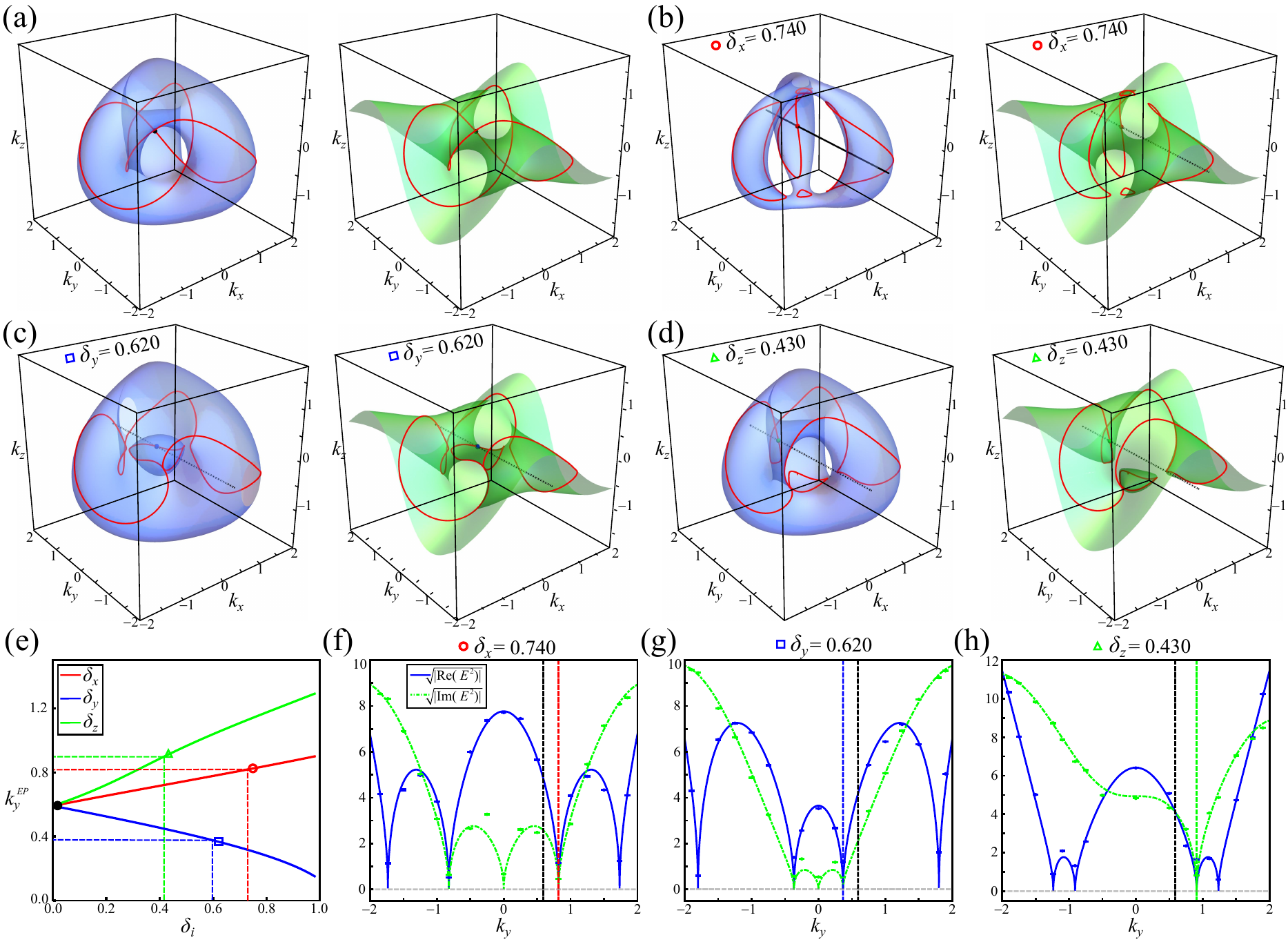}
   \caption{Robustness of symmetry-independent knotted exceptional lines versus perturbation. Blue and green surfaces respectively correspond to $\text{Re}(E^2)=0$ and $\text{Im}(E^2)=0$ for the model defined by Eq.~(7) (a), Eq.~(7) with perturbations of $\delta_x=0.740$ (b), $\delta_y=0.620$ (c) and $\delta_z=0.430$ (d) with $(p,q)=(3,2)$, $\epsilon=-20$. The solid red curves are the exceptional lines, i.e., the intersection of two surfaces. The black dot in (a) corresponds to the reference exceptional point with $\{k_x=0.338, k_y=0.591, k_z=0\}$. The solid and dot-dashed black lines in (b)-(d) indicate the parameters for (f)-(h). (e) The strength of $k_y$ of the reference exceptional point versus the perturbation of $\delta_x\sigma_x$  (red solid line), $\delta_y\sigma_y$ (blue solid line), and $\delta_z\sigma_z$ (green solid line), respectively. The critical points, where the knotted exceptional lines disappear, are $\{\delta_x=0.729, \delta_y=0.598, \delta_z=0.416\}$ as indicated by vertical dashed lines, with their corresponding $k_y^{EP}$ indicated by horizontal dashed lines. The empty circle, square and triangle correspond to the perturbation parameters for (f), (g) and (h), respectively.
   (f)-(h) The $\sqrt{|\text{Re}(E^2)|}$ (blue solid line) and $\sqrt{|\text{Im}(E^2)|}$ (green dot-dashed line) of the band structure obtained from the model defined by Eq.~(7) with different perturbations. Black and colored vertical dashed lines indicate the reference exceptional point, either without or with different perturbations. Theoretical predictions are represented by curves, and the experimental results by symbols. Error bars indicate the statistical uncertainty, obtained by assuming Poissonian statistics in the photon-number fluctuations.
}
   	\label{fig:S9}
\end{figure*}

\begin{figure*}
\includegraphics[width=\textwidth]{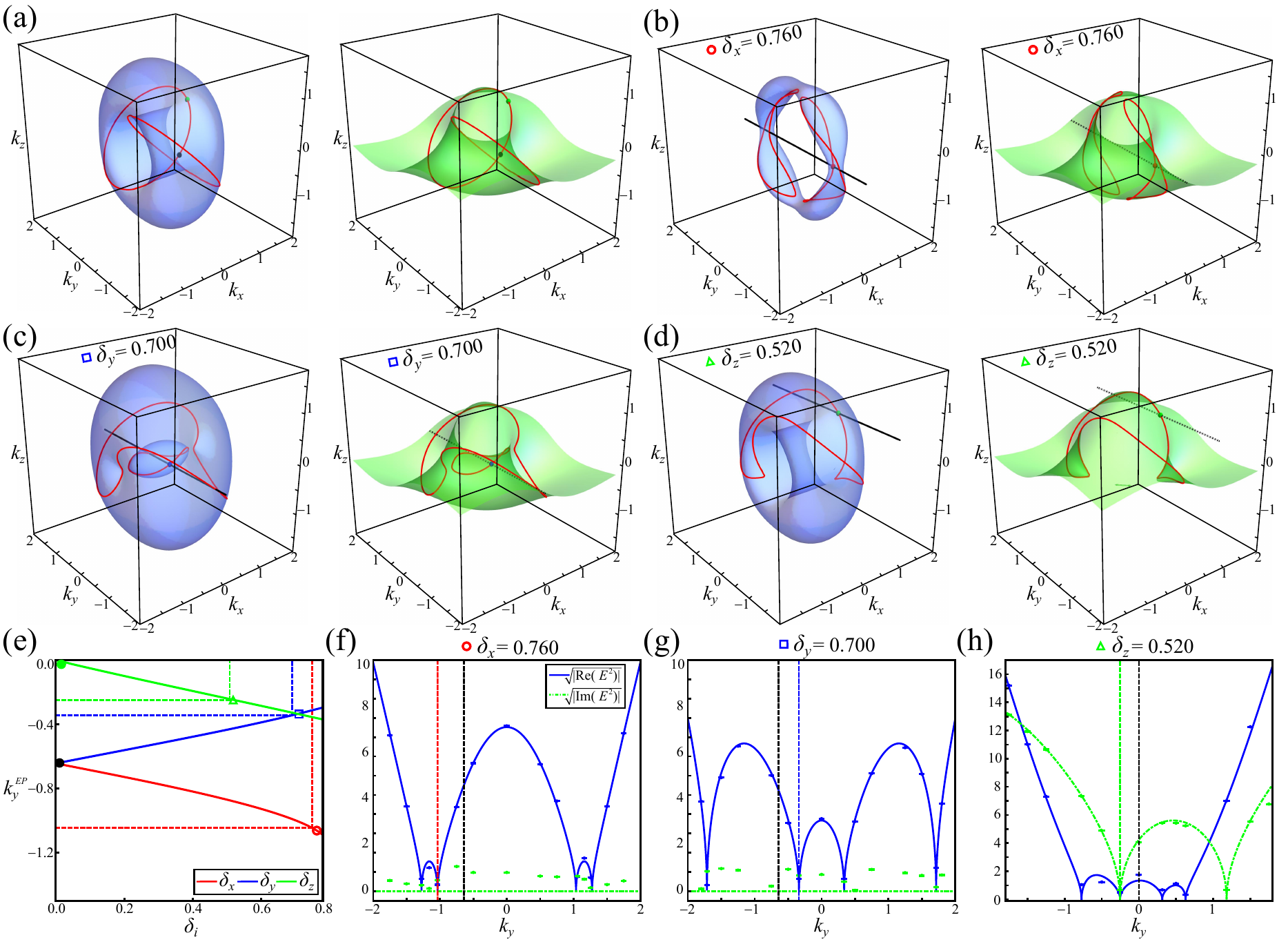}
   \caption{Robustness of symmetry-independent linked exceptional lines versus perturbation.
   Blue and green surfaces correspond to $\text{Re}(E^2)=0$ and $\text{Im}(E^2)=0$ for the model defined in Eq.~(7) (a), Eq.~(7) with perturbations $\delta_x=0.760$ (b), $\delta_y=0.700$ (c) and $\delta_z=0.520$ (d) with $(p,q)=(2,2)$, $\epsilon=-20$. The solid red curves are the exceptional lines, i.e., the intersection of two surfaces. The black and green dot in (a) correspond to the reference exceptional points with $\{k_x=0, k_y=0.338, k_z=0\}$ and $\{k_x=0.723, k_y=0, k_z=0\}$. The solid and dot-dashed black lines in (b)-(d) indicate the parameters for (f)-(h). (e) The strength of $k_y$ of the reference exceptional point versus the perturbation of $\delta_x\sigma_x$  (red solid line), $\delta_y\sigma_y$ (blue solid line), and $\delta_z\sigma_z$ (green solid line), respectively. The critical points, where the linked exceptional lines disappear, are $\{\delta_x=0.749, \delta_y=0.690, \delta_z=0.508\}$ as indicated by vertical dashed lines, with their corresponding $k_y^{EP}$ indicated by horizontal dashed lines. The empty circle, square and triangle correspond to the perturbation parameters for (f), (g) and (h), respectively. (f)-(h) The $\sqrt{|\text{Re}(E^2)|}$ (blue solid line) and $\sqrt{|\text{Im}(E^2)|}$ (green dot-dashed line) of the band structure obtained from the model defined in Eq.~(7), with different perturbations. Black and colored vertical dashed lines indicate the reference exceptional point either without or with different perturbations. Theoretical predictions are represented by curves, and the experimental results by symbols. Error bars indicate the statistical uncertainty, obtained by assuming Poissonian statistics in the photon-number fluctuations.
   }
	\label{fig:S10}
\end{figure*}

\subsection{Robustness of exceptional lines against perturbation}

We now present a quantitative analysis regarding the robustness of exceptional lines against perturbation. First, we consider the case of symmetry-protected exceptional lines, under the non-Hermitian Hamiltonian $H_1$ in Eq.~(6) of the main text. For $0\leq d_x \leq4$, the exceptional lines are robust against a small symmetry-preserving perturbation of the form $i\delta_x\sigma_x$ when $(d_x+\delta_x)^2-1/16=0$ is satisfied. The exceptional lines disappear for $\delta_x>0.25$. Under the symmetry-preserving perturbations of $\delta_y\sigma_y$ and $\delta_z\sigma_z$, the exceptional lines disappear for $\delta_y>3.99$ and $\delta_z>3.75$, respectively. In our experiment, we choose $\delta_x=0.35$, $\delta_y=4.1$ and $\delta_z=3.8$, corresponding to the red, blue and green symbols in Fig.~\ref{fig:S8}(a). As illustrated in Figs.~\ref{fig:S8}(b-d), the exceptional lines disappear in accordance with theoretical predictions.

The second case we consider is the symmetry-independent knotted exceptional lines. Under a small perturbation with the form $\delta_j\sigma_j$ ($j=x,y,z$, $0\leq\delta_j\leq 0.4$), the knotted exceptional lines are still robust. However, with sufficiently large $\delta_j$, the knotted exceptional lines disappear. We choose the exceptional point with $\{k_x=0.338,k_y=0.591,k_z=0\}$ and vary $\delta_j$. The knotted exceptional lines disappear for $\delta_x>0.729$, $\delta_y>0.598$ and $\delta_z>0.416$, respectively. In our experiment, we choose $\delta_x=0.740$, $\delta_y=0.620$ and $\delta_z=0.430$, the knotted exceptional lines disappear in accordance with theoretical predictions as shown in Fig.~\ref{fig:S9}.

Similarly, the symmetry-independent, linked exceptional lines are robust against small perturbation, but disappear when the strength of the perturbation increases, i.e., $\delta_x>0.749$, $\delta_y>0.690$ and $\delta_z>0.508$. In our experiment, we choose $\delta_x=0.760$, $\delta_y=0.700$ and $\delta_z=0.520$, the linked exceptional lines disappear in accordance with theoretical predictions as shown in Fig.~\ref{fig:S10}.

\clearpage


\end{widetext}
\end{document}